\documentstyle[12pt]{article}
\newcommand{\be}{\begin{equation}}
\newcommand{\ee}{\end{equation}}
\newcommand{\Pa}{Painlev\'e }
\newcommand{\PaP}{Painlev\'e}
\newcommand{\zz}{z-z_0}
\newcommand{\tfrac}[2]{\textstyle\frac{#1}{#2}}
\newcommand{\dfrac}[2]{\displaystyle\frac{#1}{#2}}

\begin{document}
\title{\ \vskip 0.3cm \Pa equations in terms of entire
functions\footnote{Lectures given at the summer school ``The \Pa
property, one century later'', Carg\`ese, 3-22~June, 1996}}
\author{Jarmo Hietarinta\thanks{E-mail: hietarin@newton.tfy.utu.fi}\\
Department of Physics, University of Turku\\ FIN-20014 Turku, Finland}

\maketitle

\begin{abstract}
In these lectures we discuss how the \Pa equations can be written in
terms of entire functions, and then in the Hirota bilinear (or
multilinear) form. Hirota's method, which has been so useful in
soliton theory, is reviewed and connections from soliton equations to
\Pa equations through similarity reductions are discussed from this
point of view. In the main part we discuss how singularity structure
of the solutions and formal integration of the \Pa equations can be
used to find a representation in terms of entire functions. Sometimes
the final result is a pair of Hirota bilinear equations, but for
$P_{VI}$ we need also a quadrilinear expression. The use of discrete
versions of \Pa equations is also discussed briefly. It turns out that
with discrete equations one gets better information on the
singularities, which can then be represented in terms of functions
with a simple zero.
\end{abstract}

\newpage

\section{Introduction}
Hirota's bilinear method has turned out to be very efficient in
constructing multisoliton solutions to integrable evolution equations.
But since \Pa equations do not have soliton solutions, why should we
care about writing them in the Hirota bilinear form? In these lectures
we will show that this method is relevant even for integrable ODE's.

The fundamental idea behind Hirota's direct method is the following:
\begin{itemize}\item[]
{\em Change into new variables in which the solutions have the
simplest form.}
\end{itemize}

\noindent This transformation will change at least the dependent
variable, and may sometimes be rather complicated. For solitons the
nicest possible form is the one where the soliton solution is given as
a polynomial of exponentials with exponents linear in the independent
variables (see. Sec.~2).

\Pa equations do not usually have solutions that can be written as
polynomials of exponentials, and although there are other special
solutions (rational or solutions made of special functions) to which
Hirota's method is relevant, there are other more general reason that
lead to the same.  Indeed, the idea of solutions being ``as
nice as possible'' can be extended to ODE's: We can demand that the
solutions are expressed in terms of {\em entire functions}. This is
not a new idea, it was studied by \Pa himself in
\cite{P:CR126,P:CR127}. In his Acta Mathematica paper of 1902 he
writes \cite{P:acta25}, p.14:
\begin{quote}
Puisque les int\'egrales $y(x)$ des \'equations pr\'ec\'edentes
[$P_I$, $P_{II}$, and $P_{III}$ in the present notation] sont des
fonctions {\em m\'eromorphes} dans tout le plan, il est bien
\'evident qu'elles sont repr\'esentables par le quotient de deux
fonctions {\em enti\`eres;} mais ce qu'il importe de remarquer c'est
qu'on peut choisir ces fonctions enti\`eres de mani\`ere qu'elles
v\'erifient une \'equation diff\'erentielle tr\`es simple du $3^e$
ordre.
\end{quote}
It should not be surprising that \PaP's explicit results for $P_I$,
$P_{II}$, and $P_{III}$ in \cite{P:acta25} are in the bilinear form.
More recently solutions to the \Pa equations in terms of entire
functions were considered by Lukashevich \cite{Luka70}, and in this
school K.~Okamoto will give still another method of bilinearizing
the \Pa equations.

The outline of these lectures is the following.  In Sec.~2 we will
introduce Hirota's direct (bilinear) method by discussing the soliton
solutions of the Korteweg--de~Vries equation.  In this case
``niceness'' is obvious, because the explicit soliton solutions have
the simplest possible form.  In the subsequent sections we will write
the \Pa equations in terms of entire functions, using three
methods. First in Sec.~3 we use the fact that many \Pa equations can
be obtained by similarity reductions from soliton equations with
already known bilinear forms. In Sec.~4, which is the main part, we
discuss how quadratic and quartic forms can be derived by studying the
singularity structure of the solution and then writing the equations
in terms of entire functions. (For a previous study along these lines,
see \cite{HK}.) In Sec.~5 we discuss briefly how discrete \Pa
equations (c.f.\ the talk of A.~Ramani) can be used as starting
points, because somehow the discrete formulation is more sensitive to
the singularity structure.

Finally in this introduction, let us list the equations under discussion:
\begin{eqnarray}
P_I:&&y''=6y^2+z,\label{E:PI}\\
P_{II}:&&y''=2y^3+xy+\alpha,\label{E:PII}\\
P_{III}:&&y''=\frac1y\,y'^2-\frac1z\,y'+y^3+\frac1z\,(\alpha y^2+\beta) -
\frac1y,\label{E:PIII}\\
eP_{III}:&&u''=\frac1u\,u'^2+e^{2x}u^3+e^x(\alpha u^2+\beta)-e^{2x}\frac1u,
\label{E:ePIII}\\
P_{IV}:&&y''=\frac1{2y}\,y'^2+\frac32\,y^3+4zy^2+2(z^2-\alpha)y+
\beta\, \frac1y,\label{E:PIV}\\
P_V:&&y''=\left(\frac1{2y}+\frac1{y-1}\right)\,y'^2-\frac1z\,y'\nonumber\\
&&\hskip 2cm +\alpha\frac{y(y-1)^2}{z^2}+\beta\frac{(y-1)^2}{z^2y}
+\gamma\frac yz+\delta\frac{y(y+1)}{y-1},\label{E:PV}\\
eP_V:&&u'' =\frac12\left(\frac1u+\frac1{u-1}\right)u'^2\nonumber\\
&&\hskip 1cm -\left(\alpha\frac u{u-1}+
\beta\frac{u-1}{u} +\gamma e^x u(u-1)+\delta e^{2x}u(u-1)(2u-1)\right),
\label{E:PV1e}\\
P_{VI}:&&y''=\frac12\left(\frac1y+\frac1{y-1}+\frac1{y-z}\right)y'^2
-\left(\frac1z+\frac1{z-1}+\frac1{y-z}\right)y'\nonumber\\
&&\hskip 1cm +\frac{y(y-1)(y-z)}{z^2(z-1)^2}
\left(\alpha+\beta\frac{z}{y^2}+\gamma\frac{z-1}{(y-1)^2}
+\delta\frac{z(z-1)}{(y-z)^2}\right),\label{E:PVI}\\
eP_{VI}:&&u''=\frac12\left(\frac1u+\frac1{u-1}+\frac1{u-e^x}\right)u'^2-
e^x\left(\frac1{e^x-1}+\frac1{u-e^x}\right)\nonumber\\
&&+\frac{u(u-1)(u-e^x)}{(e^x-1)^2}\left[\alpha+\frac{e^x\beta}{u^2}
+\frac{(e^x-1)\gamma}{(u-1)^2}+\frac{e^x(e^x-1)\gamma}{(u-e^x)^2}\right].
\label{E:PVIe}
\end{eqnarray}
The exponential versions are obtained by $y(z)=u(x)$ for $P_{III}$ and
$P_{VI}$, and $y(z)=\frac{u(x)}{u(x)-1}$ for $P_{V}$, where $z=e^x$,
and the primes of $u$ stand for differentiation with respect to $x$.

\section{Hirota's bilinear method for soliton equations}
Here we will briefly discuss Hirota's method \cite{Hi71} for
constructing multisoliton solutions to integrable equations (for
a review, see e.g.\ \cite{LesH,Pondi}).

\subsection{Definitions}
The first step in the construction is to transform the equation
into the Hirota form.  As an example let us consider the Korteweg --
de~Vries (KdV) equation 
\begin{equation}
u_{xxx}+6uu_x+u_t=0.
\label{E:KdV}
\end{equation}
Let us introduce the dependent variable transformation 
\begin{equation}
u=2\partial_x^2 \log F,
\label{E:tKdV}
\end{equation}
(we will see below that the new function $F$ is regular and simple for
soliton solutions) and then one can write (\ref{E:KdV}) in the
following quadratic form (after one integration):
\begin{equation}
F_{xxxx}F-4F_{xxx}F_{x}+3F^2_{xx}+F_{xt}F-F_xF_t=0.  
\end{equation} 
This does not look simpler than (\ref{E:KdV}) but
one can write it in a condensed form using the Hirota $D$ operator:
\begin{equation}
(D^4_x+D_xD_t)F\cdot F=0,
\label{E:biK}
\end{equation} 
where $D$ is a kind of antisymmetric derivative,
\begin{equation} 
D_x^nf\cdot
g=(\partial_{x_1}-\partial_{x_2})^n f(x_1)g(x_2)\big
|_{x_2=x_1=x}.\label{E:Ddef} 
\end{equation} 
The minus sign, which differentiates $D$ from Leibnitz' rule, is
crucial. We have
\[
D_x^2\,u\cdot u=uu''-u'^2,\quad D_x^4=uu''''-4u'u'''+3u''^2,
\]
etc. In the context of ODE's it is worth recalling that Borel and
Chazy arrived to these expressions by invariance theory
\cite{Cha,Bor}, and observed that they yield equations whose solutions
are entire.  For later soliton computations note that $P(D)e^{px}\cdot
e^{qx}= P(p-q)e^{(p+q)x}$.

\subsection{Multisoliton solutions}
The KdV-equation is the proto-typical representative of the class 
\begin{equation}
P(D_x,D_y,...)F\cdot F=0,\quad P(0)=0,\label{E:Cla1}
\end{equation} 
for which the multisoliton solutions are indeed simple in terms of
$F$, as opposed to $u$.  The general method of construction
multisoliton-solutions is by considering the formal expansion
\begin{equation}
F=1+\epsilon \,f_1+\epsilon^2 \,f_2+\epsilon^3 \,f_3+\cdots,
\end{equation}
where $\epsilon$ is the expansion parameter, and truncating this at
some order. The vacuum or zero-soliton solution (0SS) is given by
$F=1$.  For the one-soliton solution (1SS) only one term is needed:
it is easy to see, that due to the
antisymmetry in (\ref{E:Ddef}) $F_1=1+e^\eta$ ($\eta=p\cdot x$)
is a solution of (\ref{E:Cla1}), if the parameters $p$ satisfy a
dispersion relation $P(p)=0$. $F_1$ is the one-soliton solution (1SS)
and substitution to (\ref{E:tKdV}) yields the standard result for $u$.

The two-soliton solution (2SS) for (\ref{E:Cla1}) is obtained from the
truncation $F_2=1+f_1+f_2$, where $f_1=e^{\eta_1}+ e^{\eta_2}$.  In
order to fix $f_2$ we note that when we stay in the comoving frame of
one soliton while the other one goes to $\pm\infty$ (that is when the
other $\eta$ approaches $\pm\infty$) we should get the 1SS again.
This means that we should try
\begin{equation}
F_2=1+e^{\eta_1}+e^{\eta_2}+A_{12}e^{\eta_1+\eta_2},
\label{E:2SS}  
\end{equation} 
and substituting this  into (\ref{E:Cla1}) and using the dispersion
relation for the parameters $p_i$ we find that the equation is
satisfied, if the ``phase factor'' is given by
\begin{equation} 
A_{12}=-\frac{P(p_1-p_2)}{P(p_1+p_2)}.
\label{E:Aij}
\end{equation}  
An important point to observe is that the above works for {\em any}
polynomial $P$.  In fact there are still other classes of equations
for which the generic form has 2SS, but it should be noted that the
existence of 2SS does not imply integrability.

Although 2SS can be constructed for the whole class, 3SS work only for
certain equations, namely for the integrable ones.  It turns out that
the ansatz for a possible 3SS is fixed by the 2SS: the only one
compatible with the 2SS (\ref{E:2SS}) is
\begin{eqnarray}
F&=&1+e^{\eta_1}+e^{\eta_2}+e^{\eta_3}
+A_{12}e^{\eta_1+\eta_2}+A_{13}e^{\eta_1+\eta_3}+
	A_{23}e^{\eta_2+\eta_3}\nonumber\\
&&+A_{12}A_{13}A_{23}e^{\eta_1+\eta_2+\eta_3},
\label{E:3SS}
\end{eqnarray}
where $\eta_i=p_i\cdot x+\eta_i^0$, and the parameters $p_i$ satisfy
the dispersion relation $P(p_i)=0$ and $A_{ij}$ are given by
(\ref{E:Aij}).  This form is dictated by the requirement that as one
of the solitons goes to infinity (i.e., the corresponding $\eta$
approaches $\pm\infty$) the other two should form a 2SS (\ref{E:2SS}).

When the ansatz (\ref{E:3SS}) is substituted into (\ref{E:Cla1}) one
obtains the condition
\begin{eqnarray}&&
\sum_{\sigma_i=\pm 1}P(\sigma_1\vec p_1+\sigma_2\vec
p_2+\sigma_3\vec p_3) 
\nonumber\\ &&\ \hskip 2cm\times 
P(\sigma_1\vec p_1-\sigma_2\vec p_2)
P(\sigma_2\vec p_2-\sigma_3\vec p_3)
P(\sigma_3\vec p_3-\sigma_1\vec p_1)=0,
\label{E:3SC}
\end{eqnarray}
on the manifold defined by the dispersion relations $P(p_i)=0$.
Since the ansatz was completely fixed there
are no free coefficients, and since our principle is that
we should be able to combine {\em any} three solitons into a 3SS, we
cannot impose any new conditions on the parameters $\vec p_i$
either. Thus the condition is on the {\em equation}.

The three-soliton condition (\ref{E:3SC}) can be used to search for
integrable equations within the class (\ref{E:Cla1}) \cite{S1}.
Note that in this kind of search there are no initial assumptions
about the number of independent variables and no preferred time.
[This is in contrast with searches assuming a structure like
$u_t=F(u,u_x,u_{xx}, \dots)$.] The (nontrivial) results of this search
are as follows \cite{S1}:
\begin{eqnarray}
(D_x^4-4D_xD_t+3D_y^2)F\cdot F &=&0,\label{E:KP}\\
(D_x^3D_t+aD_x^2+D_tD_y)F\cdot F &=&0,\label{E:HSI}\\
(D_x^4-D_xD_t^3+aD_x^2+bD_xD_t+cD_t^2)F\cdot F &=&0,\label{E:my}\\
(D_x^6+5D_x^3D_t-5D_t^2+D_xDy)F\cdot F &=&0.\label{E:SKR}
\end{eqnarray}
Three of these were known before, (\ref{E:KP}) is the
Kadomtsev--Petviashvili (KP) equation, (\ref{E:HSI}) is the
Hirota--Satsuma--Ito equation, and (\ref{E:SKR}) the
Sawada--Kotera--Ramani equation.  Equation (\ref{E:my}) is the only
new equation and it is obvious that this equation could not have been
found by any ansatz assuming simple $t$ dependence. All of these
equations have also 4SS and pass the \Pa test \cite{BPT}.

Similar analysis of 2SS's and 3SS's have been performed on other types
of bilinear equations (mKdV\cite{S2}, sG\cite{S3}, nlS\cite{S4}, and
BO\cite{S4}).

\subsection{Gauge invariance and generalization to multilinearity}
We have so far discussed Hirota's method only from the
soliton point of view, but it has been found useful in other approaches as
well. In particular the $\tau$-functions (=$F$ above) have been
essential in the Kyoto school approach to integrable PDE's \cite{JM}.

One recent important observation is that Hirota forms are intimately
related to gauge invariance. It is easy to show that if $F,G,\dots$
solve some bilinear equations, so do $e^{ax}F,e^{ax}G,\dots$. But the
reverse is true as well \cite{multi}: If some quadratic expression is
gauge invariant, then all derivatives must appear as Hirota
derivatives.  The proof is simple. Consider the quadratic homogeneous
combination $ A_n(f,g):=\sum_{i=0}^n c_i\left(\partial_x^i\,f\right)
\left(\partial_x^{n-i}\,g\right).$ From the gauge invariance
$A_n(e^{ax} f,e^{ax} g)=e^{2{ax}}A_n(f,g)$ we can solve for the
constants $c_i$ and find that $ c_i=(-1)^i{N \choose i}c_0, $ so that
$A_n$ can indeed be written in terms of bilinear derivatives $D$: $
A_n(f,g)=c_0\,D_x^n f\cdot g$.

This gauge principle can be applied to higher multilinear expressions
\cite{multi}. For the cubic case one finds that for gauge invariant
expressions the derivatives appear through
\begin{equation}
T=\partial_1+j\partial_2+j^2\partial_3\, , \quad
T^*=\partial_1+j^2\partial_2+j\partial_3,
\label{E:Tdef}
\end{equation}
where the subscript indicates on which factor the derivative operates,
and $j=e^{2i\pi/3}$. The multilinear generalization is
\[
M_n^m=\sum_{k=0}^{n-1} e^{2\pi ikm/n} \partial_{k+1}, \mbox{ where }
0<m<n.
\] 

One can now search for integrable equations from the class
\begin{equation}
P(T,T^*) F\cdot F\cdot F=0,
\end{equation}
and new equations have been found in \cite{NEE95}, for example a
generalization of the KP equation
\begin{equation}
(T_x^4T_y^*+8\,T_x^3T_yT_x^*+27\,T_y^3-36\,T_x^2T_t)F\cdot F\cdot F=0,
\end{equation}
or in the non-linearized form obtained with $F=e^g$,
\begin{equation}
g_{xxxxy}+8g_{xxy}g_{xx}+4g_{xy}g_{xxx}+3g_{yyy}-4g_{xxt}=0.  
\label{E:NKP}
\end{equation}

\section{Bilinear forms and similarity reduction}
Similarity reductions of PDE's to ODE's of \Pa type are very important
theoretically, in fact the ARS conjecture \cite{ARS} states that if an
integrable PDE is reduced to an ODE, the ODE should be of \Pa type. We
will now follow this reduction path, but with a different purpose:
Since the bilinear formalism has been so useful and is well known for
soliton equations, we will use them as starting points and then apply
similarity reductions in order to derive bilinear forms for ODE's.  We
will not consider here all possible similarity reductions to \Pa
equations, but just some typical cases with direct reduction. [In many
case the connection to \Pa equations goes through rather complicated
(differential) transformations, which probably does not help in the
present objective of getting bilinear forms.]  For further references
about similarity reductions, see \cite{AC}, Sec.~6.5.15 and 7.2.

For bilinear variables the similarity reduction is always assumed to be
of the form $F(x,t)=\phi(z)e^{a(x,t)}, ...$ where the exponents are to
be determined so that the bilinear equation is in terms of $z$ only.

\subsection{$P_I$}
To get a similarity reduction to $P_I$ let us consider the KdV
equation (\ref{E:KdV}).  If one substitutes into it the ansatz \cite{AF}
\begin{equation}
u(x,t)=2t-2y(z),\quad z=x-6 t^2,
\label{E:KdVs}
\end{equation}
then one gets for $y(z)$ the equation
\begin{equation}
y'''=12yy'+1,
\end{equation}
which integrates to $P_I$ (\ref{E:PI}).

As was shown before, the bilinearization of KdV proceeds through the
dependent variable transformation (\ref{E:tKdV}) so that we now have
the relation
\begin{equation}
t-y(x-6t^2)=\partial_x^2 \log F.
\label{E:K2}
\end{equation}
This suggest that for $P_I$ we should introduce a new dependent
variable $\phi$ by
\begin{equation}
y=-(\log \phi)'',
\label{E:K3}
\end{equation}
and from (\ref{E:K2},\ref{E:K3}) we find that the similarity reduction
for the bilinear dependent variable corresponding to (\ref{E:KdVs})
should be
\begin{equation}
F=\phi(x-6t^2)e^{\frac12tx^2+a(t)x+b(t)}.
\label{E:tPI}
\end{equation}
(Note the free functions $a$ and $b$, on which we have no information
at the moment.)  When this is substituted into (\ref{E:biK}) we obtain
something that is a function of $z$ alone, if we choose $a(t)=-4t^3$
($b$ drops out). The result is then
\begin{equation}
(D_z^4+2z)\phi\cdot\phi=0,
\label{E:BPI}
\end{equation}
which is the standard bilinear form for $P_I$.  The notable feature in
the above process is the necessity of the gauge factor, in this case
$e^{\frac12tx(x-8t^2)}$.

We could also start from the Boussinesq equation
\begin{equation}
u_{xxxx}+3(u^2)_{xx}+u_{xx}-u_{tt}=0,
\label{E:Bq}
\end{equation}
and using similarity reduction $u=-2y(x-t)$ \cite{AS} we immediately
obtain $y''''=6(y^2)''$ which can be integrated twice to yield
(\ref{E:PI}) with suitable integration constants. Equation
(\ref{E:Bq}) can be bilinearized as KdV with (\ref{E:tKdV}), which
yields (after two $x$ integrations)
\begin{equation}
(D_x^4+D_x^2-D_t^2)F\cdot F=0.
\end{equation}
The similarity reduction for $F$ should now be of the form
\begin{equation}
F=\phi(x-t)e^{xa(t)+b(t)},
\end{equation}
and indeed the bilinear form (\ref{E:BPI}) follows, with $z=x-t$, if
we use the gauge $e^{-\frac12t^2x+\frac16t^3}$.

\subsection{$P_{II}$}
The second \Pa equation can be obtained by a similarity reduction from
the mKdV equation
\begin{equation}
u_{xxx}-6u^2u_x+u_t=0,
\label{E:mKdV}
\end{equation}
by the reduction ansatz \cite{AS}
\begin{equation}
u=(3t)^{-1/3}\, y(z),\quad z=x/(3t)^{1/3}.
\end{equation}
This yields for $y$ the equation
\begin{equation}
y'''=6y^2y'+zy'+y,
\end{equation}
which can be integrated to (\ref{E:PII}).

There are two ways to bilinearize mKdV.  In the conventional approach
we have to use the potential form, as for KdV.  Thus let us introduce
$v$ by $u=\partial_x v$, substitute this into (\ref{E:mKdV}) and
integrate the result with respect to $x$, this yields
\begin{equation}
v_{xxx}-2(v_x)^3+v_t=0.
\label{E:pmKdV}
\end{equation}
(The integration constant can be absorbed into $v$, since it is defined
up to an additional function of $t$.) The bilinearizing dependent variable
transformation is
\begin{equation}
v=\log\frac G F
\end{equation}
and substitution into (\ref{E:pmKdV}) yields
\begin{equation}
-FG[(D_x^3+D_t)F\cdot G]+3[(D_x^2)F\cdot G][D_xF\cdot G]=0.
\label{E:mKdVQ}
\end{equation}
At this point we have one equation for two functions, so in principle
we can introduce extra conditions for them. Recall that $F$ and $G$
are defined only up to a common multiplicative factor, so this is the
origin of the freedom we now have. For soliton solutions it turns out
that the best way to fix this factor is to demand $D_x^2 F\cdot G=0,$
then we get the bilinear form
\begin{equation}
\left\{\begin{array}{rcl}
(D_x^3+D_t)F\cdot G&=&0,\\
D_x^2 F\cdot G&=&0.
\end{array}\right.
\label{E:mKdVB2}
\end{equation}
The 1SS for this class of equations is given by $F=1+e^\eta,\,
G=1-e^\eta$ with dispersion relation given by the odd polynomial.

Maybe a general comment on equation splitting is in order here.  It
should be noted that for some other kind of solutions the above might
not be the best way to split (\ref{E:mKdVQ}).  The general method is
to put $D_x^2 F\cdot G = \lambda FG$ where $\lambda$ is an arbitrary
function, which yields the pair
\[
\left\{\begin{array}{rcl}
(D_x^3+D_t-3\lambda)F\cdot G&=&0,\\
(D_x^2-\lambda) F\cdot G&=&0.
\end{array}\right.
\]
If we now make a gauge change 
\[
F\to e^\theta F,\,G\to e^\theta G,
\]
the above equation changes to
\[
\left\{\begin{array}{rcl}
(D_x^3+D_t-3(\lambda-2\theta))F\cdot G&=&0,\\
(D_x^2-(\lambda-2\theta)) F\cdot G&=&0.
\end{array}\right.
\]
For a given type of solution (rational, soliton) one needs a specific
form of $(\lambda-2\theta)$, for soliton solutions this term should vanish.

The other bilinearization of (\ref{E:mKdV}) is obtained by
substituting $u=g/f$ directly into it, and the result can then be
split into an nlS type bilinear equation
\begin{equation}
\left\{\begin{array}{rcl}
(D_x^3+D_t)f\cdot g&=&0,\\
D_x^2f\cdot f+2g^2&=&0.
\end{array}\right.\label{E:mKS}
\end{equation}
The 1SS of this system is given by $f=1-e^{2\eta},\, g=-2pe^\eta$.

The dependent variables of these two forms
(\ref{E:mKdVB2},\ref{E:mKS}) are related by
\begin{equation}
g=D_x G\cdot F,\quad f=GF.
\end{equation}

Let us now see how the above bilinear forms can be used to bilinearize
$P_{II}$.  In the first case with bilinearization through
$u=\partial_x \log(G/F)$ the natural ansatz is
$y=\frac{d}{dz}\log\frac \psi \phi$, because then
\[
\partial_x\log\frac GF=u=\frac1{(3t)^{1/3}}y(z)=\frac1{(3t)^{1/3}}
\frac{d}{dz}\log\frac\psi\phi=\partial_x\log\frac\psi\phi,
\]
(note the partial derivatives) so that we could just try
\begin{equation}
G(x,t)=\psi(z),\,F(x,t)=\phi(z),\mbox{ with }z=x/(3t)^{1/3}.
\end{equation} 
Indeed this works, and we get from (\ref{E:mKdVB2})
\begin{equation}
\left\{\begin{array}{rcl}
(D_z^3-zD_z)\phi\cdot \psi&=&0,\\
D_z^2\phi\cdot \psi&=&0.
\end{array}\right.
\label{E:IIbil1}
\end{equation}

In the second case with $u=g/f$ 
\[
\frac{g}{f}=u=\frac1{(3t)^{1/3}}y(z)=
\frac1{(3t)^{1/3}}\frac{\Psi(z)}{\Phi(z)},
\]
suggests we should try
\begin{equation}
g=a(x,t)\Psi(z),\quad f=a(x,t)\,(3t)^{1/3}\,\Phi(z),
\end{equation}
and then  from (\ref{E:mKS}) we get a bilinear form depending only
on $z$, if we just choose $a=1$:
\begin{equation}
\left\{\begin{array}{rl}
(D_z^3-zD_z+1)\Phi\cdot \Psi=0&,\\
D_z^2\,\Phi\cdot \Phi+2\Psi^2=0&.
\end{array}\right.
\label{E:IIbil2}
\end{equation}
It is easy to check that the substitution
$\Psi=\phi\psi,\,\Phi=D_z\phi\cdot\psi$ reduces (\ref{E:IIbil2}) to
(\ref{E:IIbil1}).

\subsection{$P_{III}$}
Special cases of $P_{III}$ can be obtained by similarity reductions
\cite{AS,Oi80} from the sine-Gordon (sG) equation
\begin{equation}
u_{xt}=\sin u.
\label{E:sG}
\end{equation}
The first similarity ansatz is
\begin{equation}
u(x,t)=-i\log\,y(z),\quad z=xt,
\end{equation}
and substitution to (\ref{E:sG}) leads to the the special case
\begin{equation}
y''=\frac1y\,y'^2-\frac1z\,y'+\frac1{2z}\,(y^2-1).
\end{equation}

The sG equation (\ref{E:sG}) can be bilinearized using
\begin{equation}
u=-2i\log\frac{f+ig}{f-ig},
\end{equation}
yielding
\begin{equation}
\left\{\begin{array}{rl}
(D_xD_t-1)g\cdot f=0&,\\
D_xD_t(f\cdot f-g\cdot g)=0&.
\end{array}\right.
\label{E:sGB}
\end{equation}
The similarity reductions for $f,g$ should be
$g(x,t)=\phi(z),\,f=\psi(z)$
and they yield \cite{Oi80}
\begin{equation}
y=\left(\frac{\psi+i\phi}{\psi-i\phi}\right)^2,
\end{equation}
with
\begin{equation}
\left\{\begin{array}{rl}
(zD_z^2+\partial_z-1)\phi\cdot \psi=0&,\\
(zD_z^2+\partial_z)(\phi\cdot \phi-\psi\cdot \psi)=0&.
\end{array}\right.
\label{E:sGBs}
\end{equation}
This, however, is not satisfactory, because contains ordinary
derivatives. The trick to eliminate them is to change the dependent
variables by $\phi(z)=\bar\phi(\xi),\,\psi(z)=\bar\psi(\xi),\,z=e^\xi$,
because then we get
\begin{equation}
\left\{\begin{array}{rl}
(D_\xi^2-\xi)\bar\phi\cdot \bar\psi=0&,\\
D_\xi^2(\bar\phi\cdot \bar\phi-\bar\psi\cdot \bar\psi)=0&.
\end{array}\right.
\label{E:sGBh}
\end{equation}

Another similarity ansatz for (\ref{E:sG}) is
\begin{equation}
u(x,t)=-2i\log\,w(\zeta),\quad \zeta=2\sqrt{xt},
\end{equation}
leading to
\begin{equation}
w''=\frac1w\,w'^2-\frac1z\,w'+\tfrac14(w^3-1/w).
\end{equation}
This is related to the above as follows: If
$\bar\phi(\xi),\,\bar\psi(\xi)$ solve (\ref{E:sGBh}) then
\begin{equation}
w=\frac{\bar\psi(2\log(\zeta/2))+i\bar\phi(2\log(\zeta/2))}
{\bar\psi(2\log(\zeta/2))-i\bar\phi(2\log(\zeta/2))}.
\end{equation}
Thus we have the same basic bilinear equation (\ref{E:sGBh})
corresponding to two different nonlinear ones.

\section{Solutions in terms of entire functions}
As was mentioned before, \Pa considered already quite early the
question of representing the solutions in terms of entire functions
\cite{P:CR126,P:CR127,P:acta25}.  But how could we find such entire
functions? (\Pa does not give any constructive method, but just the
solutions.) One direct way is by studying the singularities of the
solutions and then doing some manipulations on them so that their
entireness becomes manifest \cite{HK}.

Here we would like to present an additional aspect to the introduction
of the entire functions: By choosing these functions properly one can
actually integrate the equation once.

Suppose we have an equation of the form
\begin{equation}
y''=\alpha y'^2+ \beta y'+\gamma,
\end{equation}
where $\alpha,\,\beta,\,\gamma$ are functions of $z$ and $y$, primes
stand for derivatives with respect to $z$. We want to integrate it to
the form
\begin{equation}
I:=A y'^2+B y'+C-\int_c^z D\,d\zeta,
\label{E:Ie}
\end{equation}
i.e. to find functions  $A,B,C,D$ of $z$ and $y$ such that
\begin{equation}\begin{array}{rl}
\frac{dI}{dz}\equiv & 
y''(2Ay'+B)+A_y y'^3+(A_z+B_y)y'^2+(B_z+C_y)y'+C_z-D \\
&=(2Ay'+B)(y''-\alpha y'^2-\beta y'-\gamma)=0.
\end{array}
\end{equation}
(Here the subscripts stand for partial derivatives.) This immediately
yields the set of equations
\begin{equation}\begin{array}{rcl}
A_y&=&-2\alpha A,\\
A_z+B_y&=&-2\beta A-\alpha B,\\
B_z+C_y&=&-2\gamma A-\beta B,\\
C_z-D&=&-\gamma B.
\end{array}
\label{E:inteqs}
\end{equation}

In the following, it often turns out that
\begin{equation}
f:=e^{\int\int D\, dz\, dz}
\label{E:fdef}
\end{equation}
is an entire function, and that the other entire function can be
obtained from $g:=yf$. With this definition of $f$ we get two equations
from the above:
\begin{equation}
\frac{Q_1}{f^2}\equiv (\log f)''-D(z,g/f)=0,
\label{E:fppdef}
\end{equation}
and
\begin{eqnarray}
\frac{R}{\varrho}&\equiv& (\log f)' - \int_c^z D\,dz\nonumber\\ &=& (\log
f)'- \left[A(z,\tfrac{g}{f}) (\tfrac{g}{f})'^2+B(z,\tfrac{g}{f})
(\tfrac{g}{f})'+ C(z,\tfrac{g}{f})\right]-c_1=0.
\label{E:fpdef}
\end{eqnarray}
where $c_1$ is a constant.  

Below we will show that for the \Pa equations $R$ defined above is
quartic (with $\varrho$ a simple quartic polynomial of $f$ and $g$ (no
derivatives)) and $Q_1$ quadratic in $f,g$ and their derivatives.
From these two equations further equations can be derived, including
those in Hirota bilinear form.

\subsection{$P_I$}
For $P_I$ the situation is a special and one entire function is
enough.  Using the above method we find that with $\alpha=\beta=0$ and
$\gamma=6y^2+z$ one solution to (\ref{E:inteqs}) is given by
\begin{equation}
A=\tfrac12,\quad B=0,\quad C=-(2y^3+zy),\quad D=-y.
\label{E:Iint}
\end{equation} 
\Pa mentions that $f:=e^{\int\!\int D \, dz}$ is entire when $\int
D\,dz=\tfrac12 y'^2-2y^3-yz+c_1$, in accordance with (\ref{E:Iint}).
This is easy to prove: Near any singularity the solution $y$ of
(\ref{E:PI}) behaves as \cite{HK}
\begin{equation}
y=\frac1{(\zz)^2}+O((\zz)^2),
\end{equation} 
so that at that point $f$ [as defined by (\ref{E:fdef}) with $D=-y$]
behaves smoothly:
\begin{equation}
f=(\zz)\cdot[\mbox{const }+O(\zz)].
\end{equation} 
Then from (\ref{E:fppdef}) (using $y$ in place of $g/f$) we get
\cite{P:acta25}
\begin{equation}
y=-(\log f)'',
\end{equation}
and when this is substituted into $P_I$ we get for $f$ an equation
in Hirota's bilinear form
\begin{equation}
(D_z^4+2z)f\cdot f=0.
\label{E:IB}
\end{equation}

In this paper we also want keep track of the integration constants.
Equation (\ref{E:IB}) is fourth order so it has two additional
constants of integration. They are related to the gauge invariance
under $f \to e^{\alpha+x\beta}f$, which is a common property of
equations in Hirota form (in \cite{multi} is was argued that the gauge
invariance is the {\em defining} property of Hirota form.)

In the following the final results for other \Pa equations cannot be
written as one quadratic equation but rather as a pair, so let us do
it here also. For this purpose we take another solution of
(\ref{E:inteqs}) with $A,\,C,\,D$ multiplied by 2 of what was given in
in (\ref{E:Iint}).  This yields $f=\exp(-2\int dz\int dz\, y)=
(\zz)^2\cdot[\mbox{const }+O(\zz)]$ which is needed to guarantee that
$g:=yf$ is also entire. Then from (\ref{E:fppdef})
\begin{equation}
Q_1\equiv f''f-f'^2+2fg\equiv \tfrac12D_z f\cdot f+2fg=0,
\label{E:Iq1}
\end{equation}
and from  (\ref{E:fpdef}) ($\varrho=-f^4$)
\begin{equation}
R\equiv (f'g-g'f)^2-f^3f'-4g^3f-2zgf^3-c_1f^4=0.
\label{E:I1o}
\end{equation}
These equations are equivalent to (\ref{E:PI}) in the following sense:
\begin{equation}
2(f'g-fg')P_I=Q_1+f^2\left(R/f^4\right)'.
\label{E:PQR}
\end{equation}
The pair (\ref{E:Iq1},\ref{E:I1o}) is third order, and since two
constants of integration are accounted for ($c_1$ and the overall
scale of $f,g$) only one more constant of integration remains,
and in this sense this pair represents the once integrated $P_I$.

Further equations can be derived as follows: By considering
$(R/(f^2g^2))'=0$ and using (\ref{E:Iq1}) to simplify the result we
get another quadratic equation
\begin{equation}
Q_2\equiv gg''-g'^2+ff'+zfg+c_1f^2=0,
\label{E:Iq2}
\end{equation}
which appears in \cite{Luka70}. However, it is not gauge invariant and
therefore not expressible in Hirota form, furthermore the pair
$Q_1=0,Q_2=0$ is not equivalent to (\ref{E:PI}) (one would need $R=0$ as
well). If one instead considers the combination $Q_3:=(g^2 Q_1+f^2
Q_2+R)/(fg)$ one obtains
\begin{equation}
Q_3\equiv f''g-2f'g'+fg''-zf^2-2g^2\equiv D_z^2f\cdot g-zf^2-2g^2=0,
\label{E:Iq3}
\end{equation}
and (\ref{E:PI}) is equivalent to $Q_1=0,Q_3=0$. Furthermore this pair
is in the Hirota form, and the two integration constants are related
to the gauge invariance $(f,g) \to
(e^{\alpha+x\beta}f,e^{\alpha+x\beta}g)$.

Thus in terms of the entire function $f$ and $g$ $P_I$ can be
expressed by one fourth order equation in Hirota form (\ref{E:IB})
or by the third order pair (\ref{E:Iq1},\ref{E:I1o}), or by the
pair of second order equations in Hirota form
(\ref{E:Iq1},\ref{E:Iq3}).

\subsection{$P_{II}$}
For $P_{II}$ the expansion around a singularity is \cite{HK}
\begin{equation}
y=\pm\frac1{\zz}\mp\frac{z_0}6(\zz)+\dots
\end{equation}
and if we just consider $y^2$ we get entire functions from
\begin{equation}
f:=e^{-\int\int y^2\, dz\,dz}=\zz+\dots,\quad g:=yf=\pm1+\dots
\end{equation}

The integration yields the solution
\begin{equation}
A=1,\quad B=0,\quad C=-(y^4+zy^2+2\alpha y),\quad D=-y^2.
\end{equation}
agreeing with the above.  (\Pa considers $\int D\,dz=
y'^2-y^4-zy^2-2\alpha y$ in \cite{P:CR126,P:acta25}). Then from
(\ref{E:fppdef}) we get the equation
\begin{equation}
Q_1\equiv ff''-f'^2+g^2\equiv \tfrac12 D_z^2 f\cdot f+gg=0,
\label{E:IIq1}
\end{equation}
and from (\ref{E:fpdef})  ($\varrho=-f^4$)
\begin{equation}
R\equiv (f'g-g'f)^2-f^3f'-g^4-zg^2f^2-2\alpha g f^3-c_1f^4=0.
\label{E:IIr}
\end{equation}
(both given by \Pa in \cite{P:acta25}). Equation (\ref{E:PQR}) holds
also for $P_{II}$.

As before, another quadratic equation \cite{Luka70} is obtained from
$(R/(f^2g^2))'=0$
\begin{equation}
Q_2\equiv gg''-g'^2+ff'+\alpha gf+c_1f^2=0.
\label{E:IIq2}
\end{equation}
(Note that here $z$ is absent.) Instead of this one could consider the
gauge invariant (and $c_1$ independent) expression $Q_3:=(g^2 Q_1+f^2
Q_2+R)/(fg)$, i.e.,
\begin{equation}
Q_3\equiv f''g-2f'g'+fg''-\alpha f^2-zfg\equiv (D_z^2-z)f\cdot g-\alpha
f^2=0.
\label{E:IIq3}
\end{equation}
The Hirota bilinear pair (\ref{E:IIq1},\ref{E:IIq3}) is the same as
given in \cite{HK}. The counting of integration constants is as before.

Still another form is obtained if we take $f=FG,\, g=D_z F\cdot G$
corresponding to $y=F'/F-G'/G$ \cite{P:acta25}, which leads to the
bilinear form
\begin{equation}
\left\{\begin{array}{rcl}
D_z^2\,F\cdot G &=&0,\\
(D_z^3-zD_z-\alpha)\,F\cdot G&=&0.
\end{array}\right.
\label{E:bilP2}
\end{equation}
This is fifth order and there are 3 obvious integration constants
related to the invariance under $F\to ae^{cx} F,\, G\to be^{cx} G$.

\subsection{$P_{III}$}
For $P_{III}$ the expansion around a movable singularity is
\begin{equation}
y=\pm\frac1{\zz}-\frac{\alpha\pm 1}{2z_0}+\dots
\end{equation}
and if one consider the combination $z(y^2+\frac{\alpha}{z}y)
=z_0/(\zz)^2+ O(1)$ on finds that
\begin{equation}
f:=e^{-\int\frac{dz}z\int z(y^2+\frac{\alpha}{z}y)dz},\quad g:=yf
\end{equation}
are entire  \cite{HK}.

The term $\frac{dz}z$ above suggests that it might be better to work
with the exponential version (\ref{E:ePIII}) (as was done by \PaP,
\cite{P:acta25}). Then one solution to the integration problem is
\begin{equation}\begin{array}{rcl}
\bar A=1/u^2,\quad \bar B=0,\quad
\bar C&=&2e^x(\beta/u-\alpha u)-e^{2x}(1/u^2+u^2),\\
\bar D&=&2e^x(\beta/u-\alpha u)-2e^{2x}(1/u^2+u^2).
\end{array}
\end{equation}
This corresponds to \PaP's $2\zeta$ in \cite{P:acta25} p.15. However
this does not directly lead to entire functions, and \Pa adds some ad
hoc operations, which in fact amount to using another solution
\begin{equation}\begin{array}{rcl} 
A=1/(4u^2),\, B=-1/(2u),\,
C&=&\frac12(e^x(\beta/u-\alpha u)-\frac12e^{2x}(1/u^2+u^2)),\\ 
D&=&-(e^x\alpha u+e^{2x}u^2).
\end{array}
\end{equation}
This leads directly to the desired result:
$f:=e^{\int\int D\, d^2x}$ and $g:=uf$ are entire, and we get
\begin{equation}
Q_1\equiv f''f-f'^2+\alpha e^x fg+ e^{2x}g^2
\equiv \tfrac12D_x^2f\cdot f+\alpha e^x fg+e^{2x}g^2=0,
\label{E:IIIq1}
\end{equation}
and ($\varrho=-4f^2g^2$)
\begin{equation}
R\equiv (f'g-g'f)^2-2fg(f'g+g'f)+f^2g^2-2e^x(\alpha fg^3-\beta f^3g)-
e^{2x}(g^4+f^4)-4c_1f^2g^2.
\label{E:IIIr}
\end{equation}
As usual, by considering $(R/(f^2g^2))'=0$ we get another equation,
which now happens to be in the Hirota form:
\begin{equation}
Q_2\equiv gg''-g'^2-\beta e^x fg-\delta e^{2x} f^2
\equiv \tfrac12 D_x^2 g\cdot g-\beta e^x fg+e^{2x} f^2=0 .
\label{E:IIIq2}
\end{equation}
Thus $P_{III}$ is equivalent either to the pair
(\ref{E:IIIq1},\ref{E:IIIr}) or (\ref{E:IIIq1},\ref{E:IIIq2}),
in the first case the system is third order and there are two
integration constants, the overall scale and $c_1$, in the second case
the system is fourth order with two-parameter gauge freedom. 

Note that $P_{III}$ is invariant under $u\to 1/u$ accompanied with the
parameter changes $\alpha \to -\beta,\,\beta\to -\alpha$. This
corresponds to $f\leftrightarrow g$ and we see from the above that
it is indeed a symmetry of the bilinear equations.  Thus it might be
said that the zeroes of $u$ are as important singularities as its
poles, and to handle all of them at the same time one could define
functions $F,G,K,M$ by \cite{P:acta25}
\begin{equation}
f=FG,\quad g=KM,\quad e^xu=\frac{G'}G-\frac{F'}F,\quad
\frac{e^x}u=\frac{M'}M-\frac{K'}K.
\end{equation}
In that case we get four equations for four entire functions, two from
the above definitions
\begin{equation}
D_x\,G\cdot F=e^xKM,\quad D_x\,M\cdot
K=e^xFG,
\label{E:P3b21}
\end{equation}
and two from $Q_1,Q_2$
\begin{equation}
D_x^2\,F\cdot G=-\alpha e^xKM,\quad D_x^2\,M\cdot K=\beta e^xFG.
\label{E:P3b22}
\end{equation}
The system is now 6th order with four scale related integration
constants: $F\to ab e^{cx} F,\, G\to a/be^{cx}G,\, K=\to km e^{cx} K,
M=\to k/m e^{cx} M$. The result (\ref{E:P3b21},\ref{E:P3b22}) is more
symmetric, but involves twice as many dependent variables. Whether
it is more useful in practical applications depends on the problem.

\subsection{$P_{IV}$}
For $P_{IV}$ the expansion reads
\begin{equation}
y=\frac{\pm1}{\zz}-z_0+\dots
\end{equation}
and one finds that 
\begin{equation}
f:=e^{-\int dz\int dz (y^2+2zy)},\quad g:=yf,
\label{E:P4e}
\end{equation}
define entire functions \cite{HK}. 

The integration method again works with the simplest choice. If we use
the solution
\begin{equation}
A=\frac1{4y},\quad B=0,\quad C=-\tfrac14(y^3+4zy^2+4(z^2-\alpha)y
+2\beta/y),\quad D=-(y^2+2zy),
\end{equation}
agreeing with (\ref{E:P4e}), we get the equations
\begin{equation}
Q_1\equiv ff''-f'^2+g^2+2zfg
\equiv \tfrac12 D_z^2 f\cdot f+g^2+2zfg=0,
\label{E:IVq1}
\end{equation}
and ($\varrho=-4f^3g$)
\begin{equation}
R\equiv (f'g-g'f)^2-4f^2gf'-g^4-4zfg^3
-4(z^2-\alpha)f^2g^2+2\beta f^4-2c_1f^3g=0,
\label{E:IVr}
\end{equation}
and the other quadratic equation is
\begin{equation}
Q_2\equiv g''g-g'^2+2gf'-2\beta f^2+c_1fg=0.
\label{E:IVq2}
\end{equation}
The gauge invariant combination of the above turns out to be
trilinear:
\begin{equation}
T\equiv T_zT_z^* f\cdot g\cdot g-2(\beta
f^3+2(z^2-\alpha)fg^2+zg^3)=0.
\label{E:IVt}
\end{equation}
($T_zT_z^* f\cdot g\cdot g=gD_z\,g\cdot f+\frac12fD_z\,g\cdot g$.) One
can now show that $P_{IV}$ can be expressed as a linear combination of
either $Q_1$ and $(R/(f^3g))'$ or $Q_1$ and $T$, with the usual
accounting of integration constants.

At this point we would like to return to the question of gauge
trasformations, briefly mentioned before. The point is that the
function
\[
f:=e^{-\int dz\int dz (y^2+2zy)+p(z)},
\]
is entire for any fixed polynomial $p$ of $z$. For example note that
$P_{IV}$ has the polynomial solution $y=-\frac23 z$, and then from
$(\log f)''=-y^2-2zy$ we would get $f\propto e^{\frac8{27}z^4}$. It
would clearly be desirable to have polynomial $f,g$ as well. This
could be obtained by a proper choice of $p$, see \cite{Gromak}, p.~68
for details. The same problem exists for $P_{III}$, see \cite{Gromak},
p.~90.

\subsection{$P_{V}$}
In this case again the nicest results are obtained for a specific form of
the equation. Computations with the standard form reveal that one
should instead consider the equation (\ref{E:PV1e}) obtained from the
standard one by $y(z)=u(x)/(u(x)-1)$, $z=e^x$ \cite{HK}.
The expansion for $u$ is given by
\begin{equation}
u=\pm\frac{i/\sqrt{2\delta}}{\zz}-\frac{\pm i\sqrt{2\delta}+
\gamma-2\delta z_0}{4\delta z_0}+\dots
\end{equation}
and using the method of \cite{HK} leads one to the entire functions
\begin{equation}
f:=e^{\int dx\int dx(\gamma e^x u+2\delta e^{2x}u(u-1)},\quad g:=uf.
\end{equation}

The integration has the corresponding solution
\begin{equation}
\begin{array}{rcl}
A=\dfrac1{2u(u-1)},\quad B=0,\quad C&=&-\alpha\dfrac 1{u-1}-
\beta\dfrac1{u} +\gamma e^x u+\delta e^{2x}u(u-1),\\
D& =&\gamma e^x u+2\delta e^{2x}u(u-1),
\end{array}
\end{equation}
and from (\ref{E:fppdef}) we get the first equation 
\begin{equation}
Q_1\equiv
\tfrac12D_x^2f\cdot f-\gamma e^x fg-2\delta e^{2x}g(f-g)=0, 
\end{equation} 
and from (\ref{E:fpdef}) ($\varrho=-2f^2g(g-f)$)
\begin{eqnarray}
R&\equiv&(f'g-fg')^2-2fg(g-f)f'-c_1f^2g(g-f)\nonumber\\
&&\quad-2f^3[\alpha g+\beta(g-f)] 
+2(g-f)g^2[\gamma e^x f-\delta e^{2x} (g-f)]=0.
\end{eqnarray}
From the derivative $(R/(f^2g^2))'=0$ one obtains
equation
\begin{equation}
Q_2\equiv g''g-g'^2-gf'-2\beta f^2+(\alpha+\beta-c_1/2)fg=0,
\end{equation}
and the gauge invariant linear combination of the above is again
trilinear,
\begin{equation}
T\equiv T_xT_x^* (f-\tfrac23g)\cdot g \cdot g -2\alpha fg^2
-2\beta f(f-g)^2-\gamma e^x g^2(2f-g)-2\delta e^{2x} g^2(f-g)=0.
\end{equation}
($T_xT_x^* (f-\tfrac23g)\cdot g \cdot g=gD_x^2f\cdot
g+(f/2-g)D_x^2g\cdot g$)  Furthermore one finds that $P_{V}$ is
expressible as a linear combination of $Q_1$ and $(R/(f^2g(f-g)))'$ or
of $Q_1$ and $T$.

\subsection{$P_{VI}$}
For $P_{VI}$ the situation is more complicated. In fact the method
used in \cite{HK} does not work: there are no polynomials of $u$ alone
from which entire functions can be built. On the other hand \Pa in
\cite{Pa06} proposes an expression that is supposed to yield an entire
function, but this expression involves also $u'$. Presumably one can
search such expressions using the expansion around the singularity,
but we will here take a different route.

It turns out that Lukashevich \cite{Luka70} obtained some
quadratic and quartic expressions for $P_{VI}$ as well, but we could
not verify the precise forms given in \cite{Luka70}. Using these
results as a guide we searched for two quadratic and a quartic
expression with similar properties as before, using (\ref{E:PVIe}). 
This resulted in
\begin{equation}
Q_1:=(e^x-1)^2(f''f-f'^2)+(e^x-1)fg'+2\alpha g(g-f)-
(\alpha+c_1)(e^x-1)fg,
\end{equation}
\begin{equation}
Q_2:=e^{-x}(e^x-1)^2(g''g-g'^2)+(e^x-1)f'g+\beta e^xf(g-f)
-(\beta-\delta-\gamma+c_1)(e^x-1)fg,
\end{equation}
and
\begin{eqnarray}
R&:=&(e^x-1)^2(f'g-fg')^2-2(e^x-1)fg(f-g)(e^xf'-g')\nonumber
\\&&-2\alpha g^2(f-g)(e^xf-g)+2\beta e^x f^2(f-g)(e^xf-g)
-2\gamma(e^x-1) f^2g(e^xf-g\nonumber)\\&&-2\delta e^x(e^x-1)f^2g(f-g)
+2c_1(e^x-1)fg(f-g)(e^xf-g).
\label{E:P6R}
\end{eqnarray}
The relationships between these expressions and the $P_{VI}$ equation
are as follows:
\begin{eqnarray}
&&2(e^x-1)^2f^3g(f-g)(e^xf-g)P_{VI}\\
&&\hskip 2.5cm=2(-g^2Q_1+e^xf^2Q_2)(f-g)(e^xf-g)+(e^xf^2-g^2)R,\nonumber
\end{eqnarray}
and
\begin{equation}
M\,P_{VI}=2e^x(Q_1-Q_2)+(e^xf^2-g^2)(e^x-1)^2(R/U)',
\end{equation}
where
\begin{eqnarray*}
U&:=&(e^x-1)fg(f-g)(e^xf-g),\\
M&:=&-2(e^x-1)^3f^2\left[(e^x-1)(f'g-fg')(e^xf^2-g^2)-e^xfg(f-g)^2\right]/U
\end{eqnarray*}
(Note a spurious solution: $M$ vanishes if $u$ solves
$(e^x-1)u'(u^2-e^x)=e^xu(u-1)^2$.)

Finally we have a relation between $Q_1,\,Q_2$ and $R$ as
\begin{equation}
BQ_1+CQ_2+(e^x-1)fgV(R/V)'=0,
\end{equation}
where
\begin{eqnarray*}
B&:=&-2g^2\left[(e^x-1)(f'g-fg')-e^xf(f-g)\right],\\
C&:=&2f^2\left[e^x(e^x-1)(f'g-fg')-e^xf(f-g)\right],\\
V&:=&(e^x-1)^2f^2g^2,
\end{eqnarray*}

As far as  gauge invariant expressions are concerned, one finds that
\[ X Q_1+Y Q_2+Z R\]
is gauge independent whenever
\[
X+Y=2Z(f-g)(e^xf-g),
\]
and then the $c_1$ terms vanish as well. One possibility is to take
\begin{eqnarray}
Q_1-Q_2&\equiv&\tfrac12(e^x-1)^2(D_x^2\,f\cdot f-e^{-x}D_x^2\,g\cdot g)
+(e^x-1)D_x\,g\cdot f\nonumber\\&&+2(\alpha g-\beta e^x f)(g-f)
-(\alpha-\beta+\gamma+\delta)fg(e^x-1)=0,
\label{E:P61}
\end{eqnarray}
which is bilinear. For the other expression we could not get any
simplification so it is quadrilinear, for example:
\begin{eqnarray}
&&(e^x-1)^2(f-g)(e^xf-g)D_x^2 f\cdot f\nonumber\\
&&\quad +(e^x-1)^2(D_x f\cdot g)^2
-2e^x(e^x-1)f(f-g)(D_z f\cdot g)\nonumber\\
&&\quad -2\alpha g(f-g)(e^xf-g)((e^x-1)f-g)
+2\beta e^xf^2(f-g)(e^xf-g)\nonumber\\
&&\quad -2\gamma (e^x-1)f^2g(e^xf-g)
-2\delta e^x(e^x-1)f^2g(f-g)=0.
\label{E:P62}
\end{eqnarray}

Let us now return to the integration procedure and try to understand
why the straightforward procedure failed. Substituting $g=uf$ into
(\ref{E:P6R}) shows that
\begin{equation}
\frac{R}{-2(e^x-1)fg(f-g)(e^xf-g)}=(\log f)'-[Au'^2+Bu'+C]-c_1,
\end{equation}
where
\begin{equation}
A=\frac{e^x-1}{2u(u-1)(u-e^x)},
\quad B=\frac{-1}{u-e^x},\quad C=-\frac{\alpha u}{e^x-1}+
\frac{e^x\beta}{(e^x-1)u}+\frac{\gamma}{u-1}+\frac{e^x\delta}{u-e^x}.
\end{equation}
This expression is in fact in \cite{Pa06}, (eq.\ (3), $m=Au'^2+Bu'+C$
from above), and \Pa states that $e^{\int m}$ has no singularities,
apart from the fixed ones (for $z$ they are at 0,1 and $\infty$, for
$x(=\log z)$ at $-\infty,\,0,\,\infty$).

The integration procedure therefore works as before up to this point,
and the problem is in $D$ of (\ref{E:Ie}), it is no longer a function
of $u$ only.  Indeed, if one writes again
\begin{equation}
I:=Au'^2+Bu'+C-\Delta
\end{equation}
and identifies $\Delta$ with $(\log f)'$ then from $Q_1$ we get
\begin{equation}
(e^x-1)\Delta'+u(\Delta-\alpha -c_1)+\frac{2\alpha u(u-1)}{e^x-1}=0.
\end{equation}
(Similar expressions were considered in \cite{Gar}.)

For $P_{VI}$ the situation has then turned out to be quite different from
the others, as might have been expected. Nevertheless, even in this
case the final result can be written in multilinear form, in this
case we need one bilinear (\ref{E:P61}) and one quadrilinear
(\ref{E:P62}) expression.

\section{Discrete \Pa}
At the moment the most interesting developments in the field of
integrable systems seem to take place in the area of integrable {\em
difference} equations. Most properties of continuous integrable
systems can be extended to the discrete case, e.g., Lax pairs,
existence of solitons (for partial difference equations) and the \Pa
test. [For an overview see the lectures of Nijhoff and Ramani.]

In order to define discrete \Pa equations one should have a definition
of discrete \Pa property. Grammaticos, Ramani and Papageorgiou
\cite{GRP} have proposed that {\em singularity confinement} is the
proper discrete analogue of the \Pa property. Singularity confinement
means that if a mapping leads to singularity, then after a finite
number of steps one should get again out of it and this should take
place without essential loss of information. Singularity confinement
has subsequently been used to generate discrete forms of \Pa
equations, in fact several families of them. (Before calling some
difference equation a discrete version of a differential equation, one
must verify at least that its continuum limit is the original
continuous equation, but the continuous and discrete equations should
share some other properties as well.)

The bilinear approach has a natural analogue in the discrete case. It
is best stated using the gauge principle: If the expression is
homogeneous in the dependent variables $F,G,\dots$ and invariant under
$F(n)\to F(n)e^{pn},\,G(n)\to G(n)e^{pn},\dots$ then we say it is in
Hirota form.

As an example let us consider d-$P_I$\cite{ORGT}. One version is
\begin{equation}
\overline w + w + \underline w=\frac{z}w+a,
\label{E:D2}
\end{equation}
where $\overline w=w(n+1),\,w=w(n),\underline w=w(n-1)$ and $z=\alpha n
+\beta$. If in this equation one hits a singularity, it is by first
arriving somehow to $w(k)=0$, and a closer study indicates that the
sequence of special $w$ values are $\{0,\infty,\infty,0\}$, after
which regular values are again obtained. This pattern of $w$ values is
obtained from
\begin{equation}
w(n)=\frac{F(n+2)F(n-1)}{F(n+1)F(n)},
\label{E:D1}
\end{equation}
if $F$ has a simple zero $F(k-1)=0$. The expression (\ref{E:D1}) is
homogeneous and gauge invariant, and if one substitutes it to the
discrete derivative of (\ref{E:D2}) one arrives to
\begin{eqnarray}
&&F(n+3)F(n-1)F(n-2)-F(n+2)F(n+1)F(n-3)=\nonumber\\
&&\hskip 2cm z(n)F(n+1)^2F(n-2)-z(n-1)F(n+2)F(n-1)^2.
\label{E:dp1}
\end{eqnarray}
This is the trilinear version of d-$P_I$.

If one now applies the continuous limit to the above, the previous results
are obtained \cite{ORGT}: If $a=6$, $w=1+\epsilon^2 y,\,
z=-3+\epsilon^4\zeta,\, \zeta=n\epsilon$, one finds that $y=2(\log
F)_{\zeta\zeta}$ and (\ref{E:dp1}) becomes the $z$ derivative of
(\ref{E:IB}) divided by $f^2$.

The most important aspect of the above is the way the complicated
singularity pattern of $w$ is obtained from a simple zero of $F$. This
is the discrete analogue of expressing the original solution in terms
of entire functions. In the discrete case the process is much more
clear and this is an indication that discrete systems are more
fundamental.

If the system has several singularity patterns we need more functions
to handle them, in general the number of singularity patterns is the
same as the number of entire functions. In \cite{ORGT} this idea was
followed to its logical conclusion: for d-$P_{VI}$ the authors obtained
a set of bilinear equations involving 8 functions. We will not repeat
all of their results here, just some illustrative examples.

One version of d-$P_{II}$ is 
\begin{equation}
\overline w+\underline w=\frac{zw+a}{1-w^2}.
\label{E:d3}
\end{equation}
It has two singularity patterns, $\{-1,\infty,1\}$ and
$\{1,\infty,-1\}$. The entrance to the first pattern and exit from the
second is described by
\begin{equation}
w(n)=-1+\frac{F(n+1)G(n-1)}{F(n)G(n)}
\end{equation}
while for the remaining part we would get
\begin{equation}
w(n)=-1-\frac{F(n-1)G(n+1)}{F(n)G(n)}.
\end{equation}
Equating these two expressions we get the first equation
\begin{equation}
F(n+1)G(n-1)+F(n-1)G(n+1)-2F(n)G(n)=0,
\end{equation}
whose continuous limit yields the first equation of (\ref{E:bilP2}).
The second equation is obtained from (\ref{E:d3}). In this case we
have two singularity patterns and functions, and the structure of the
patterns determines one equation between the functions.

For the higher discrete \Pa equations the singularity patterns
sometimes determine everything. More precisely, the singularity
patterns suggest different ways of writing the dependent variable in
terms of functions with simple zeroes, and comparing these expressions
one gets enough equations. As an example let us consider d-$P_{III}$
\begin{equation}
\overline w\underline w=\frac{cd(w-az)(w-bz)}{(w-c)(w-d)}
\end{equation}
where $z=\lambda^n$ (corresponding to change of variables $z=e^x$ in
the continuous form) the singularity patterns are $\{c,\infty,d\}$,
$\{d,\infty,c\}$, $\{az,0,bz\}$, and $\{bz,0,az\}$. This suggest the
representations \cite{ORGT}
\begin{equation}
w=c\left(1-\frac{F(n+1)G(n-1)}{F(n)G(n)}\right)
=d\left(1-\frac{F(n-1)G(n+1)}{F(n)G(n)}\right)=\frac{HK}{FG},
\end{equation}\begin{equation}
\frac1w=\frac1{az}\left(1-\frac{H(n+1)K(n-1)}{H(n)K(n)}\right)=
\frac1{bz}\left(1-\frac{H(n-1)K(n+1)}{H(n)K(n)}\right)=\frac{FG}{HK}.
\end{equation}
Equating these expressions and taking suitable sums and differences
yields four bilinear equations in a nice symmetric form
\begin{eqnarray*}
F(n+1)G(n-1)+F(n-1)G(n+1)-2F(n)G(n)&=&-\left(\frac1c+\frac1d\right)H(n)K(n),\\
F(n+1)G(n-1)-F(n-1)G(n+1)&=&\left(\frac1d-\frac1c\right)H(n)K(n),\\
H(n+1)K(n-1)+H(n-1)K(n+1)-2H(n)K(n)&=&-z(a+b)F(n)G(n),\\
H(n+1)K(n-1)-H(n-1)K(n+1)&=&z(b-a)F(n)G(n).
\end{eqnarray*}
The continuous limit is obtained with $z=e^{2\epsilon n}=e^{2x} ,\,
a=\epsilon-a_0\epsilon^2,\, b=-\epsilon-b_0\epsilon^2,\,
c=1/\epsilon+c_0,\, d=-1/\epsilon+d_0$ and yields 
\begin{eqnarray*}
D_x  F\cdot G&=&-HK,\\
D_x  H\cdot K&=&-e^{2x}FG,\\
D_x ^2 F\cdot G&=&(c_0+d_0)HK,\\
D_x ^2 H\cdot K&=&(a_0+b_0)e^{2x}FG.
\end{eqnarray*}
One can now verify, that if one uses substitution $u=-e^{-x}
\frac{d}{dx}\log(F/G)$ in (\ref{E:ePIII}) the result vanishes due to
the above equations.

The important point in the above construction is that the singularity
patterns determined {\em all} of the bilinear equations. The \Pa equation is
then just a way to represent the singularity patterns through one
function and its equation. This situation continues with some of the
higher discrete \Pa equations.

The approach of taking the discrete singularity patterns seriously and
using them to derive bilinear forms \cite{ORGT} is systematic and
powerful. In this way the Hirota bilinearization of $P_{VI}$ was first
obtained. The drawback is in the proliferation of dependent variables:
for $P_{VI}$ 8 functions are needed, one for each singularity type
(the singularities are at $y=0,\infty,1,$ and $z$, with the next term
in the expansion having $\pm$ sign). In the bilinear+quadrilinear form
(\ref{E:P61},\ref{E:P62}) we manage with two functions, and this is
enough also for Okamoto's method (truly bilinear but using also
ordinary derivatives and therefore not gauge invariant). Which form is
best will then depend on the practical problem on hand, and it is
useful to keep all alternatives in mind.

\section*{Acknowledgments}
I would like to thank M. Kruskal for discussions and B. Grammaticos
for comments on the manuscript. This work was supported in part by the
Academy of Finland, project 31445.

\end{document}